\documentstyle[12pt]{article}

\begin{document}

{\sf \begin{center}
\noindent
{\Large \bf Non-Riemannian effective spacetime effects on Hawking radiation in superfluids}\\[3mm]

by \\[0.3cm]

{\sl L.C. Garcia de Andrade}\\ 

\vspace{0.5cm}
Departamento de F\'{\i}sica
Te\'orica -- IF -- Universidade do Estado do Rio de Janeiro-UERJ\\[-3mm] 
Rua S\~ao Francisco Xavier, 524\\[-3mm]
Cep 20550-003, Maracan\~a, Rio de Janeiro, RJ, Brasil\\[-3mm]
Electronic mail address: garcia@dft.if.uerj.br\\[-3mm]

\vspace{2cm}
{\bf Abstract}
\end{center}
\paragraph*{}

Riemannian effective spacetime description of Hawking radiation in $^{3}He-A$ superfluids is extended to non-Riemannian effective spacetime. An example is given of non-Riemannian effective geometry of the rotational motion of the superfluid vacuum around the vortex where the effective spacetime Cartan torsion can be associated to the Hawking giving rise to a physical interpretation of effective torsion recently introduced in the literature in the form of an acoustic torsion in superfluid $^{4}He$ (PRD-70(2004),064004). Curvature and torsion singularities of this $^{3}He-A$ fermionic superfluid are investigated. This  Lense-Thirring effective metric , representing the superfluid vacuum in rotational motion, is shown not support Hawking radiation when the isotropic $^{4}He$ is restored at far distances from the vortex axis. Hawking radiation can be expressed also in topological solitons (moving domain walls) in fermionic superfluids in non-Riemannian (teleparallel) $(1+1)$ dimensional effective spacetime. A teleparallel solution is proposed where the quasiparticle speed is determined from the teleparallel geometry.
\vspace{0.5cm}
\noindent
{\bf PACS numbers:} \hfill\parbox[t]{13.5cm}{02.40, 0450} 

\newpage
\section{Introduction}

\paragraph*{}

The investigation of the non-Riemannian geometry of vortex acoustics \cite{1} has been proved to be a powerful tool to explain certain features of the rotation in fluid analog models that have not been taken into account when use is made of the Riemannian geometry only  \cite{2}. In this paper we provide a physical explanation for these sort of effective spacetime torsion by computing the Hawking radiation temperature spectrum \cite{3} of a rotational superfluid vacuum $^{3}He-A$ around the vortex leading to a Doppler shift of the quasiparticle energy \cite{4}, in the context of the non-Riemannian effective spacetime geometry. It is important to stress  that on a previous paper \cite{1} we have consider the acoustic non-Riemannian geometry applied to $^{4}He$ superfluid while here we consider the above type of fermionic superfluid \cite{4} investigating what happens.  Besides we point out that the Lense Thirring model in teleparallel gravity  has been investigated by Pereira et al \cite{5}, where the axial torsion vector is proportional to the angular momentum of the Kerr general relativistic metric. Here we extend their case to a more general Riemann-Cartan effective spacetime for superfluids. Rotating cylinders in the context of Einstein-Cartan gravity have also recently been presented \cite{6}. We also show that in case of moving domain wall in $(1+1)$ dimensional effective teleparallel spacetime, the only nonvanishing component coincides with the surface gravity and consequently with Hawking temperature.  Other type of non-Riemannian effective spacetime, called Finsler spacetime  has been investigated Visser et al \cite{7}. The paper is organised as follows: In section II we compute using the method of Cartan calculus of differential forms the curvature and torsion of the rotational superfluid vacuum and express the torsion components in terms of the effective gravity and Hawking temperature at the horizon. In the third section, as another example, we consider the extension of Hawking radiation from Riemannian to non-Riemannian spacetime in moving domain wall in $^{3}He-A$, where a teleparallel model is proposed. Finally in section IV we present some discussions and future prospects.
\section{Non-Riemannian geometry in superfluid rotating vacuum and Hawking radiation}

\paragraph*{}
In this section we shall present the non-Riemannian geometry of superfluid vacuum effective metric similar to the Lense-Thirring (1918) \cite{8} spacetime in general relativity. The rotational metric representing a quantized vortex in $^{3}He$ is described by the line element
\begin{equation}
ds^{2}= -dt^{2}+\frac{r^{2}}{c^{2}(r)}(d{\phi}-{\omega}_{LT}dt)^{2}+{c^{-2}}dr^{2}+{c_{||}}^{-2}dz^{2}
\end{equation}
where $c(r)$ and $c_{||}$ and $c_{\perp}$ are the asymmetric  "speeds of light" in the anisotropic superfluid. Here ${\omega}_{LT}(r)= \frac{{\kappa}}{2{\pi}r^{2}}$. To understand better the nature of Hawking radiation here we need to express the metric in the form where the horizon is more transparent \cite{4}. In the case of $^{4}He$ this form is given by \cite{4} 
\begin{equation}
ds^{2}= -(1-\frac{{\Omega}^{2}r^{2}}{c^{2}})dt^{2}+2\frac{{\Omega}}{c^{2}(r)}d{\phi}dt+{c^{-2}}d{\vec{r}^{2}}
\end{equation}
where the Lense-Thirring angular velocity \cite{10} ${\omega}_{LT}= -{\Omega}$ and the angular velocity is given by the expression $\vec{v}_{S}=-\vec{\Omega}{\times}\vec{r}$, $\vec{v}_{S}$ being the superfluid flow speed. The second term in the represents the frame dragging in Einstein curved spacetime of general relativity. Let us now compute the Riemann-Cartan curvature tensor and Cartan torsion tensor of the effective non-Riemannian spacetime from Cartan calculus of exterior differential forms \cite{9}. To express the metric of the quantum vortex in $^{3}He-A$ in terms of the line element 
\begin{equation}
ds^{2}= -({\omega}^{0})^{2}+({\omega}^{1})^{2}+({\omega}^{2})^{2}+({\omega}^{3})^{2}
\end{equation}
we write down the basis one-form ${\omega}^{a}$ ,$(a=0,1,2,3)$ as 
\begin{equation}
{\omega}^{0}=dt
\end{equation}
\begin{equation}
{\omega}^{1}= \frac{r}{c}(d{\phi}-{\omega}_{LT}dt) 
\end{equation}
\begin{equation}
{\omega}^{2}=\frac{dr}{c}
\end{equation}
\begin{equation}
{\omega}^{3}=\frac{dz}{c_{||}}
\end{equation}
Computing the exterior derivatives of the basis one-form one obtains the only non-vanishing derivative as
\begin{equation}
d{\omega}{^2}= \frac{dr}{c}{\wedge}d{\phi}-3{\omega}_{LT}dr{\wedge}dt-\frac{c'r}{c}dr{\wedge}(d{\phi}-{\omega}_{LT}dt)
\end{equation}
where the slash represents the radial derivative of the quasiperticle speed (phonon in the case of $^{4}He$. From Cartan's structure equations 
\begin{equation}
{R^{a}}_{b}:={R^{a}}_{bcd}({\Gamma}){\omega}^{c}{\wedge}{\omega}^{d}=d{{\omega}^{a}}_{b}+{{\omega}^{a}}_{e}{\wedge}{{\omega}^{e}}_{b}
\end{equation}
\begin{equation}
{T^{a}}=d{{\omega}^{a}}+{{\omega}^{a}}_{e}{\wedge}{{\omega}^{e}}
\end{equation}
where ${\Gamma}$ represents the Riemann-Cartan connection, ${{\omega}^{a}}_{b}$ is the conection one-form ,while ${R^{a}}_{bcd}$ is the Riemann-Cartan tensor and  ${R^{a}}_{b}$ is the Riemann-Cartan curvature two-form, and $T^{a}$ the torsion two-form, we obtain the nonvanishing components of the one-form connection 
\begin{equation}
{{\omega}^{1}}_{0}= -{[1-c'r]}{r{\omega}_{LT}}{d{\phi}}
\end{equation}
\begin{equation}
{{\omega}^{1}}_{2}= -{[1-c'r]}{d{\phi}}
\end{equation}
\begin{equation}
{{\omega}^{2}}_{0}= {[3-c'r]}{{\omega}_{LT}}{dr}
\end{equation}
while the torsion two-form is given by  
\begin{equation}
T^{1}= -{[1-c'r]}{r{\omega}_{LT}}{d{\phi}}{\wedge}dt-{[1-c'r]}{d{\phi}}{\wedge}\frac{dz}{c_{||}}
\end{equation}
which allows us to write the two only nonvanishing components of the torsion tensor as
\begin{equation}
{T^{1}}_{t{\phi}}= -{[1-c'r]}{r{\omega}_{LT}}
\end{equation}
\begin{equation}
{T^{1}}_{z{\phi}}= -\frac{[1-c'r]}{c_{||}}
\end{equation}
where we have made use of the following expression for the torsion two-form
\begin{equation}
T^{a}= {T^{a}}_{bc}{\omega}^{b}{\wedge}{\omega}^{c}
\end{equation}
where ${T^{a}}_{bc}$ is the Cartan torsion tensor \cite{8}. Since in the case of isotropic ${4}^He$ superfluid the three "light speeds" in the superfluid coincide as $c_{\perp}=c_{||}=c(r)$ , and $c'(r)=$ the torsion expressions simplify to 
\begin{equation}
{T^{1}}_{t{\phi}}= -{r{\omega}_{LT}}
\end{equation}
\begin{equation}
{T^{1}}_{z{\phi}}= -\frac{1}{c}
\end{equation}
Thus one may say that result of reference \cite{1} that the vorticity of the superfluid is proportional to torsion in $^{4}He$ superfluid is confirmed here. To a better physical understanding of the role of torsion in the realm of superfluids is better understood by performing the ratio of both torsion components
\begin{equation}
[{\frac{{T^{1}}_{t{\phi}}}{{T^{1}}_{z{\phi}}}}]_{^{4}He} = \frac{r{\omega}_{LT}}{c}
\end{equation}     
Note that the term $r{\omega}_{LT}= {v}_{S}$ one note that the order of the ratio is proportional to the "relativistic" ratio $\frac{v_{S}}{c}$ between the superfluid flow speed and the "light" speed c. Since we know that the Hawking radiation is given by
\begin{equation}
{T_{H}} = \frac{h}{2{\pi}{\kappa}_{B}}{k}
\end{equation} 
where $k=\frac{dc}{dr}$ is the surface gravity defined at the event horizon where $v_{S}=c$. Thus from this expressions and the expressions from torsion one obtains
\begin{equation}
{T_{H}} = \frac{2{\pi}{\kappa}_{B}}{h}\frac{d{T^{1}}_{r{\phi}}}{dr}
\end{equation} 
Therefore we must conclude that the temperature spectrum comes from variations of torsion components. Now from the curvature Cartan equation
\begin{equation}
{R^{1}}_{2}= d{{\omega}^{1}}_{2}+{{\omega}^{1}}_{0}{\wedge}{{\omega}^{0}}_{1}
\end{equation}
\begin{equation}
{R^{1}}_{0}=d{{\omega}^{1}}_{0}+{{\omega}^{1}}_{2}{\wedge}{{\omega}^{2}}_{0}
\end{equation}
\begin{equation}
{R^{2}}_{0}=d{{\omega}^{2}}_{0}
\end{equation}
These equations in turn allows us to obtain the expressions for the Riemann-Cartan curvature components
\begin{equation}
({R^{1}}_{2r{\phi}})_{^{3}He}= (c'+c" r)+{{\omega}_{LT}}^{2}r\frac{(1-c'r)(3-c'r)}{c}
\end{equation}
\begin{equation}
({R^{1}}_{0r{\phi}})_{^{3}He}= -[\frac{1}{c}(c-2cc'r+cc"r^{2}+3-4c'r+{c'}^{2}r^{2}){\omega}_{LT}+(1-c'r){{{\omega}'}_{LT}}r^{2}]
\end{equation}
These quite cumbersome expressions in $^{3}He-A$ fermionic superfluid become very simple in $^{4}He$ superfluid where $c'$ vanishes and the curvature expressions are reduced to
\begin{equation}
({R^{1}}_{2r{\phi}})_{^{4}He}= \frac{{{\omega}_{LT}}^{2}r}{c}
\end{equation}
\begin{equation}
({R^{1}}_{0r{\phi}})_{^{4}He}= -[\frac{(c+3)}{c}{\omega}_{LT}+{{{\omega}'}_{LT}}r^{2}]
\end{equation}
Note that the first curvature expression is proportional, since c is constant, to the Lense-Thirring acceleration. An interesting observation is that the teleparallel geometry \cite{10} where all components of Riemann-Cartan curvature tensor ${R^{a}}_{bcd}$ vanishes, is not compatible with the superfluid $^{4}He$ in this model since from the last expressions one would obtain the solution ${\omega}_{LT}=0$, which means that the superfluid rotation would vanish, and the effective spacetime would be trivially Minkowskian. However, the teleparallel geometry would fit quite well in $^{3}He-A$ superfluids. The teleparallel constraint would imply that the quasiparticle speed would be determined from the geometry. 
\section{Hawking radiation in teleparallel effective moving domain wall spacetime?} 
In this section we extend the idea of Jacobson of Volovik \cite{11} of investigating the Hawking radiation topological solitons of the order parameter (moving domain wall) in Riemannian effective geometry to non-Riemannian effective spacetime where Hawking radiation as in the previous example is related to effective spacetime torsion. In the comoving coordinate system of domain wall we can consider the effective spacetime metric in $(1+1)$ dimensions in the form \cite{4}
\begin{equation}
ds^{2}=-dt^{2}+\frac{1}{c^{2}(x)}(dx-v_{S}dt)^{2}
\end{equation}
As in previous section we make use of the Cartan calculus to compute the curvature and torsion components. The one-form basis this time are
\begin{equation}
{\omega}^{0}= dt
\end{equation}
\begin{equation}
{\omega}^{1}= \frac{1}{c(x)}(dx-v_{S}dt)
\end{equation}
This computation is much more trivial and straightforward that the Lense-Thirring superfluid example which lead us to the following only nonvanishing component of connection one-form
\begin{equation}
{{\omega}^{1}}_{0}=\frac{c'}{c}v_{S}{\omega}^{1}
\end{equation}
This expression in turn allows us to compute the effective spacetime torsion and curvature for the moving domain wall 
\begin{equation}
{T^{1}}_{10}= \frac{c'}{c}v_{S}
\end{equation}
\begin{equation}
{R^{0}}_{1xt}=\frac{{v^{2}}_{S}}{c^{2}}[2c'-c"]    
\end{equation}
The expression for torsion allows us to immeadiatly identify torsion with the surface gravity since at the black hole event horizon $v_{S}=c$ and the Hawking temperature in this case is 
\begin{equation}
T_{H}= \frac{h}{2{\pi}k_{B}}{T^{1}}_{10}
\end{equation}
this result gives a nice physical interpretation for torsion in terms of Hawking temperature spectrum which lacks in the previous paper on non-Riemannian acoustic geometry \cite{1}. The curvature expression would gives us a simple solution for the quasiparticle spee of "light" in the teleparallel case where ${R^{0}}_{1xt}=0$. This would imply a differential equation for the speed c like $2c'-c"=0$ which would give $c(x)=de^{(\frac{x}{d})}$ where d is the thickness of the domain wall. Although this does not coincide with the Jacobson-Volovik ansatz for such a soliton
\begin{equation}
c^{y}(x)=c_{\perp} , c^{x}(x)=-c_{\perp}tanh\frac{x}{d}
\end{equation}
it is interesting to point it out that the Jacobson-Volovik solution is a combination of the teleparallel solution $c(x)=de^{(\frac{x}{d})}$. This actually would be expected since most teleparallel theories of gravity would not add many physical new features but simply express the gravitational theory, namely general relativity in the context of torsion theories of gravity.   
\section{Conclusions}
A physical interpretation of effective torsion previously introduced in the literature in terms of acoustic torsion is proposed based on Hawking radiation. It is shown that in both examples of fermionic and bosonic superfluids the Hawking radiation is given a new geometrical interpretation in terms of effective geometry of spacetime torsion. The teleparallel geometry is shown to have some difficulties to fit into the $^{4}He$ superfluids but is able to fit reasonable well within the framework of fermionic $^{3}He-A$. Future prospects include to examine these ideas in the context of other physical analog models with effective torsion such as Bose-Einstein condensates and spinning string gravity. 
\section*{Acknowledgements}
\paragraph*{}
I am very much indebt to Professor Bill Unruh for his constant invaluable advice on acoustic geometry of fluids. Grants from CNPq (Ministry of Science of Brazilian Government) and Universidade do Estado do Rio de Janeiro (UERJ) are acknowledged.

\end{document}